\documentclass[a4paper,reqno,12pt,final]{amsart}
\usepackage{amsmath,amsfonts,amssymb,amsthm,amstext,array}
\usepackage[latin1]{inputenc}
\DeclareMathOperator{\AdS}{AdS}
\DeclareMathOperator{\dvol}{dvol}
\renewcommand{\d}{\partial}
\newcommand{\RR}{\mathbb{R}}
\newcommand{\EE}{\mathbb{E}}
\newcommand{\fso}{\mathfrak{so}}
\newcommand{\fh}{\mathfrak{h}}
\newcommand{\half}{\tfrac{1}{2}}
\begin{document}

\title{Penrose limits and maximal supersymmetry}
\author[Blau]{Matthias Blau}
\address{The Abdus Salam ICTP, Trieste, Italy}
\email{mblau@ictp.trieste.it}
\author[Figueroa-O'Farrill]{José Figueroa-O'Farrill}
\address{Department of Mathematics, University of Edinburgh, UK}
\email{j.m.figueroa@ed.ac.uk}
\author[Hull]{Christopher Hull}
\address{Department of Physics, Queen Mary, University of London, UK}
\email{c.m.hull@qmul.ac.uk}
\author[Papadopoulos]{George Papadopoulos}
\address{Department of Mathematics, King's College, London, UK}
\email{gpapas@mth.kcl.ac.uk}
\thanks{EMPG-02-01, QMUL-PH-02-01}
\begin{abstract}
  We show that the maximally supersymmetric pp-wave of IIB superstring
  and M-theories can be obtained as a Penrose limit of the
  supersymmetric $\AdS\times S$ solutions.  In addition we find that
  in a certain large tension limit, the geometry seen by a brane probe
  in an $\AdS\times S$ background is either Minkowski space or a
  maximally supersymmetric pp-wave.
\end{abstract}
\maketitle
\tableofcontents

\section{Introduction}

It has recently been shown \cite{NewIIB} that ten-dimensional type IIB
supergravity admits a maximally supersymmetric pp-wave background
analogous to the one discovered by Kowalski-Glikman \cite{KG} for
eleven-dimensional supergravity and discussed more recently in
\cite{FOPflux}.  In both cases, the geometry is given by a lorentzian
symmetric space $G/K$ with solvable $G$, and the field strengths (the
self-dual five-form in type IIB and the four-form in eleven
dimensions) are parallel and null; such solutions were called
Hpp-waves in \cite{FOPflux}.  The spacetime is geodesically complete
and the metric is that of a pp-wave, but the transverse geometry is
not asymptotically flat. The existence of these Hpp-wave solutions is
a little puzzling. The reason is that they are to be treated on the
same footing as the other maximally supersymmetric solutions: flat
space and solutions of the form $\AdS \times S$; but whereas these
latter solutions play the role of asymptotic or near horizon limits of
fundamental brane solutions, no such role was known for the former.
This puzzle motivated the present work, in which we will show that
these maximally supersymmetric Hpp-waves are obtained as \emph{Penrose
  limits} of the maximally supersymmetric $\AdS\times S$ solutions of
eleven-dimensional and type IIB supergravity theories.  In addition,
we shall show that the geometry seen by a brane probe in an $\AdS
\times S$ spacetime in the weak coupling (equivalently \emph{large
  tension}) limit is either a Minkowski space or a maximally
supersymmetric Hpp-wave.

In \cite{PenrosePlaneWave} Penrose showed that every spacetime has a
limit which a neighbourhood of a null geodesic becomes a pp-wave
spacetime.  Although his paper focused on the case of four-dimensional
spacetimes, he pointed out that this persists in higher dimensions.
More recently, in \cite{GuevenPlaneWave} Güven extended Penrose's work
to supergravity backgrounds in ten and eleven dimensions. He achieved
this by extending the limiting procedure to the other fields present
in the supergravity theories. (This limit had already appeared in the
context of world-sheet sigma-model actions and `contractions' of WZW models
in \cite{SfetsosPL,ORSPL,SfetsosTseytlinPL,TseytlinExact}). The limiting
solutions are characterised by having a pp-wave-like geometry and null
fluxes.  We will review this briefly in Section~\ref{sec:PLimits}.

In Section~\ref{sec:NHPL} we exhibit the maximally supersymmetric
Hpp-waves \cite{KG,FOPflux,NewIIB} of eleven-dimensional and IIB
supergravity as Penrose limits of the maximally supersymmetric
$\AdS\times S$ solutions.  Similar methods also yield the Hpp-wave
solutions to five- and six-dimensional supergravities obtained
recently in \cite{Meessen}, as well as Hpp-wave solutions in
supergravity theories in various dimensions.  These will appear in a
forthcoming paper \cite{Limits} where we present a systematic and
comprehensive discussion of the Penrose limit in string theory.  We
also show that the Penrose limit provides a natural explanation for
the structure of the group of isometries of the maximally
supersymmetric Hpp-waves.

In Section~\ref{sec:dynamics} we explore the worldvolume dynamics of a
brane probe in a spacetime under the Penrose limit. There are
different Penrose limits for each spacetime which depend on the choice
of null geodesic. The effect of the limit is to blow up a
neighbourhood of the geodesic. The Penrose limit also blows up the
induced worldvolume of branes. This can be seen by placing a brane
probe in a spacetime. A Penrose limit can be taken in which the brane
tension is scaled and at the same time a spacetime coordinate
transformation is performed.  There is a limit in which the tension of
the probe goes to infinity and the spacetime in the neighbourhood of a
null geodesic goes to the Penrose limit.

\section{Penrose limit of supergravity theories}
\label{sec:PLimits}

In this section we will briefly review the Penrose limit as described
by Güven for backgrounds of ten- and eleven-dimensional
supergravities.

Let $(M,g)$ be a lorentzian spacetime.  According to
\cite{PenrosePlaneWave,GuevenPlaneWave} in a neighbourhood of a
segment of a null geodesic $\gamma$ containing no conjugate points, it
is possible to introduce local coordinates $U,V,Y^i$ such that the
metric takes the form
\begin{equation}
  \label{eq:metric}
  g = dV \left(dU + \alpha dV + \sum_i \beta_i dY^i\right) +
  \sum_{i,j} C_{ij} dY^i dY^j~,
\end{equation}
where $\alpha$, $\beta_i$ and $C_{ij}$ are functions of all the
coordinates, and where $C_{ij}$ is a symmetric positive-definite
matrix.  The coordinate system breaks down as soon as $\det C = 0$,
signalling the existence of a conjugate point.  The coordinate $U$ is
the affine parameter along a congruence of null geodesics labelled by
$V$ and $Y^i$.  The geodesic $\gamma$ is the one for which $V=0=Y^i$.

In ten- and eleven-dimensional supergravity theories there are other
fields besides the metric, such as the dilaton $\Phi$, gauge
potentials or more generally $p$-form potentials $A_p$ with
$(p+1)$-form field strengths.  The gauge potentials are defined up to
gauge transformations $A_p \mapsto A_p + d\Lambda_{p-1}$ in such a way
that the field strength $F_{p+1} = dA_p$ is gauge invariant.  It is
possible to use this gauge freedom in order to gauge away some of the
components of the $p$-form potentials.  Indeed, one can choose a gauge
locally in which
\begin{equation}
  \label{eq:gauge}
  i(\d/\d U) A = 0~,
\end{equation}
or in components
\begin{equation*}
  A_{Ui_1i_2\dots i_{p-1}} = A_{UVi_1i_2\dots i_{p-2}} = 0~.
\end{equation*}
Similar results apply for field strengths with intereaction terms,
$F=dA+\dots $. The starting point of the Penrose limit is the data
$(M,g,\Phi,A_p)$ defined in a neighbourhood of a conjugate-point-free
segment of a null geodesic $\gamma$ where $g$ and $A_p$ take the forms
\eqref{eq:metric} and \eqref{eq:gauge}, respectively.

We now introduce a positive real constant $\Omega >0$ and rescale the
coordinates as follows
\begin{equation}
  \label{eq:diffeo}
  U = u~, \qquad V = \Omega^2 v \qquad\text{and}\qquad Y^i =
\Omega y^i~.
\end{equation}
Acting with this diffeomorphism on the tensor fields of the theory we
obtain an $\Omega$-dependent family of fields $g(\Omega)$,
$\Phi(\Omega)$ and $A_p(\Omega)$.  The coordinate and gauge choices
\eqref{eq:metric} and \eqref{eq:gauge} ensure that the following
\emph{Penrose limit} \cite{PenrosePlaneWave} (as extended by Güven
\cite{GuevenPlaneWave} to fields other than the metric) is
well-defined:
\begin{equation}
  \label{eq:rescaling}
  \begin{aligned}[m]
    \bar g &= \lim_{\Omega\to 0} \Omega^{-2} g(\Omega)\\
    \bar \Phi &= \lim_{\Omega\to 0} \Phi(\Omega)\\
    \bar A_p &= \lim_{\Omega\to 0} \Omega^{-p} A_p(\Omega)~.
  \end{aligned}
\end{equation}

By virtue of \eqref{eq:diffeo} the limiting fields only depend on the
coordinate $u$, which is the affine parameter along the null geodesic.
The resulting expression for the metric is of the form
\begin{equation}
  \label{eq:PL}
  \bar g = du dv + \sum_{i,j} \bar C_{ij}(u) dy^i dy^j~.
\end{equation}
The gauge potentials $\bar A_p$ only have components in the
transverse directions $y^i$,
\begin{equation*}
  i(\d/\d u) \bar A_p = 0 =  i(\d/\d v) \bar A_p~,
\end{equation*}
and the field strengths $\bar F_{p+1}$ are therefore of the form
\begin{equation}
  \label{eq:PLFS}
  \bar F_{p+1} = du \wedge \bar A_p(u)'~,
\end{equation}
where ${}'$ denotes $d/du$.  Note that $\bar F_{p+1}$ is null.

As the supergravity actions transform homogeneously
\cite{GuevenPlaneWave} under the scaling \eqref{eq:rescaling}, $(\bar
g, \bar \Phi, \bar A_p)$ will be a solution to the supergravity
equations of motion whenever $(g,\Phi,A_p)$ is. These and other
hereditary properties (in the sense of \cite{Geroch}) of Penrose
limits will be discussed in detail in \cite{Limits}.

The above expression for $\bar g$ is that of a pp-wave in Rosen
coordinates.  It is possible to change to Brinkman (also called
harmonic) coordinates in such a way that the resulting metric takes
the form
\begin{equation}
  \label{eq:CWtype}
  \bar g = 2 dx^+ dx^- + \left( \sum_{i,j} A_{ij}(x^-) x^i x^j \right)
   (dx^-)^2 + \sum_i dx^i dx^i~.
\end{equation}
When $A_{ij}$ is constant this metric describes a lorentzian symmetric
Cahen--Wallach space \cite{CahenWallach}.  Such spaces include the
maximally supersymmetric Hpp-waves of eleven-dimensional
\cite{KG,FOPflux} and IIB supergravity \cite{NewIIB}, namely
\begin{align}
  \label{eq:11dHpp}
  g_{11} &= 2 dx^+ dx^- - \left( \sum_{i,j=1}^3 \delta_{ij} x^i x^j
    + \tfrac14\sum_{i,j=4}^9 \delta_{ij} x^i x^j \right) (dx^-)^2 +
  \sum_{i=1}^9 dx^i dx^i\\
  \label{eq:IIBHpp}
  g_{\text{IIB}} &= 2 dx^+ dx^- -  \sum_{i,j=1}^8 \delta_{ij} x^i
  x^j  (dx^-)^2 + \sum_{i=1}^8 dx^i dx^i
\end{align}
(up to an overall scaling of $A_{ij}$ by a real positive constant
which can always be absorbed into a scaling of $(x^+,x^-)$).  It is
this observation which gives rise to the investigation reported here
and in \cite{Limits}.

The explicit change of variables which takes the metric from Rosen to
Brinkman form is given by
\begin{equation*}
  u = 2 x^- \qquad v = x^+ - \half \sum_{i,j} M_{ij}(x^-) x^i x^j \qquad
  y^i = \sum_{j} Q^i_j(x^-) x^j~,
\end{equation*}
where $Q^i_j$ is an invertible matrix satisfying (a ${}'$ now denotes
$d/dx^-$)
\begin{equation}
  \label{eq:q}
  C_{ij}Q^i_k Q^j_l = \delta_{kl} \qquad\text{and}\qquad
  C_{ij} \left(Q'{}^i_j Q^j_l - Q'{}^i_k Q^j_l\right) = 0~,
\end{equation}
and
\begin{equation*}
  M_{ij} = C_{kl}Q'{}^k_i Q^l_j~,
\end{equation*}
which is symmetric by virtue of the second equation in \eqref{eq:q}.
This equation guarantees that the limiting metric $\bar g$ has the
form \eqref{eq:CWtype}. The relation between $C_{ij}$ and $A_{ij}$ is
\begin{equation*}
  A_{ij} = -[C_{kl}Q'{}^l_j]'Q^k_i~.
\end{equation*}

It is possible to rewrite the field strengths $\bar F_{p+1}$ given in
\eqref{eq:PLFS} in terms of Brinkman coordinates, to arrive at the
following expression:
\begin{multline*}
  \bar F_{p+1} = \sum_{i_k,j_k} \frac{d}{dx^-} \bar A_{i_1i_2\dots
    i_p}(2x^-) Q^{i_1}_{j_1} Q^{i_2}_{j_2} \cdots Q^{i_p}_{j_p} \\
  \times dx^- \wedge dx^{j_1} \wedge dx^{j_2} \wedge \cdots \wedge
  dx^{j_p}~.
\end{multline*}

\section{Penrose limit of $\AdS\times S$ solutions}
\label{sec:NHPL}

In this section we exhibit the maximally supersymmetric Hpp-wave
solutions to eleven-dimensional and IIB supergravity as Penrose limits
of $\AdS\times S$ supergravity solutions.

\subsection{The metrics}
\label{sec:metrics}

The near horizon geometry of the M2-, M5- and D3-brane solutions is of
the form $\AdS_{p+2} \times S^{D-p-2}$ where the values of $p$ and $D$
corresponding to each of the above branes are listed in
Table~\ref{tab:pDradii} along with the ratio $\rho :=
R_{\AdS_{p+2}}/R_{S^{D-p-2}}$ of the radii of curvature of the two
factors.

\begin{table}[h!]
  \begin{center}
    \setlength{\extrarowheight}{5pt}
    \begin{tabular}{|c|c|c|c|}
      \hline
      Brane & $p$ & $D$ & $\rho$\\[3pt]
      \hline
      M2 & 2 & 11 & $\half$ \\
      D3 & 3 & 10 & 1 \\
      M5 & 5 & 11 & 2\\
      \hline
    \end{tabular}
    \vspace{8pt}
    \caption{Dimensions and radii of curvature}
    \label{tab:pDradii}
  \end{center}
\end{table}

The metric for anti-de~Sitter space $\AdS_{p+2}$ with radius of
curvature $R_{\AdS}$ can be written as
\begin{equation}
  \label{eq:ads}
  g_{\AdS} = R_{\AdS}^2 \left[-d\tau^2 + (\sin\tau)^2 \left(
  \frac{dr^2}{1+r^2} + r^2 d\Omega^2_{p}\right)\right]~,
\end{equation}
where $d\Omega^2_{p}$ is the $p$-sphere metric.

Similarly, we write the round metric on the $n$-sphere $S^n$ with
radius of curvature $R_S$ as
\begin{equation}
  \label{eq:sphere}
  g_S = R_S^2 \left[d\psi^2 + (\sin\psi)^2 d\Omega^2_{n-1}\right]~,
\end{equation}
where $d\Omega^2_{n-1}$ is the metric on the equatorial ($n-1$)-sphere
and $\psi$ is the colatitude.

The metric on $\AdS_{p+2} \times S^{D-p-2}$ is then $g = g_{\AdS} +
g_S$, which is given by
\begin{multline*}
  R^{-2} g = \rho^2 \left[-d\tau^2 + (\sin\tau)^2 \left(
  \frac{dr^2}{1+r^2} + r^2 d\Omega^2_{p}\right)\right]\\
   + d\psi^2 + (\sin\psi)^2 d\Omega^2_{D-p-3}~,
\end{multline*}
where we have introduced the ratio $\rho$ defined above and where $R$
is the radius of curvature of the sphere.  Let us now change
coordinates in the $(\psi,\tau)$ plane to
\begin{equation}
  \label{eq:lightcone}
  u = \psi +\rho \tau \qquad  v = \psi -\rho \tau~,
\end{equation}
in terms of which, the metric $g$ becomes
\begin{multline*}
  R^{-2} g = du dv + \rho^2 \sin((u-v)/2\rho)^2 \left(
  \frac{dr^2}{1+r^2} + r^2 d\Omega^2_{p}\right)\\ + \sin((u+v)/2)^2
  d\Omega^2_{D-p-3}~.
\end{multline*}
We now take the Penrose limit along the null geodesic parametrised by
$u$.  In practice this consists in dropping the dependence on
coordinates other than $u$.  Doing so we find
\begin{equation}
  \label{eq:PLRosen}
  R^{-2} \bar g = du dv + \rho^2 \sin(u/2\rho)^2
  ds^2(\EE^{p+1}) + (\sin (u/2))^2 ds^2(\EE^{D-p-3})~,
\end{equation}
which is the metric of a Cahen--Wallach symmetric space in Rosen
coordinates (compare with \cite{CKG} for the $d=11$ solution).

To see this let us introduce coordinates $y^a$ for $a=1,\dots,D-2$ in
such a way that the metric \eqref{eq:PLRosen} becomes
\begin{equation*}
  R^{-2} \bar g = du dv + \sum_{a=1}^{D-2} \frac{(\sin(\lambda_a
  u))^2}{(2\lambda_a)^2} dy^a dy^a~,
\end{equation*}
where
\begin{equation}
  \label{eq:lambdas}
  \lambda_a =
  \begin{cases}
    1/2\rho & a=1,\dots,p+1\\
    1/2 & a=p+2,\dots,D-2~.
  \end{cases}
\end{equation}
We change coordinates to $(x^+, x^-, x^a)$ where
\begin{equation}
  \label{eq:BrinkmanCoords}
  x^- = u/2~, \quad x^+ = v - \tfrac14 \sum_a y^a y^a
  \frac{\sin(2\lambda_a u)}{2\lambda_a}~, \quad x^a= y^a
  \frac{\sin(\lambda_a u)}{2\lambda_a}~,
\end{equation}
in such a way that the metric now becomes
\begin{equation}
  \label{eq:PLCW}
  R^{-2} \bar g = 2 dx^+ dx^- - 4\left( \sum_a \lambda_a^2 x^a x^a
  \right) (dx^-)^2   + \sum_a dx^a dx^a~,
\end{equation}
which we recognise as a Cahen--Wallach metric \eqref{eq:CWtype} whose
matrix $A_{ij}$ is constant and diagonal with negative eigenvalues
$\{-\lambda_a^2\}$.  For $\lambda_a$ given as in \eqref{eq:lambdas} we
obtain, if $\rho=\half$ or $\rho=2$, precisely the metrics of the
maximally supersymmetric Hpp-waves of eleven-dimensional supergravity
\eqref{eq:11dHpp} discovered in \cite{KG} (see also \cite{FOPflux}),
and if $\rho=1$ the maximally supersymmetric Hpp-wave of IIB
supergravity \eqref{eq:IIBHpp} discovered in \cite{NewIIB}.  The two
cases $\rho=\half$ and $\rho=2$ are isometric---an explicit
diffeomorphism being given by
\begin{equation*}
  \begin{aligned}[m]
    x^-&\mapsto \half x^-\\
    x^+&\mapsto 2x^+\\
    (x^1,\dots,x^6,x^7,\dots,x^9)&\mapsto
    (x^4,\cdots,x^9,x^1,\dots,x^3)~.
  \end{aligned}
\end{equation*}

\subsection{The $p$-forms}

The near horizon geometries of the M2, D3 and M5 brane solutions carry
fluxes with respect to ($D-p-2$)-form field strengths.  Let us
consider the M-branes first.  The $7$-form in the $\AdS_4\times S^7$
solution is given by the Hodge dual of the $4$-form
\begin{equation*}
  F_4 = \sqrt{6 |s|} \dvol(\AdS_4)~,
\end{equation*}
whereas the $4$-form in the near horizon geometry of the M5 brane
is given by
\begin{equation*}
  F_4 = \sqrt{6 |s|} \dvol(S^4)~,
\end{equation*}
where $s$ is the scalar curvature of the supergravity solution.  The
scalar curvature of the solution is $1/8$ of the scalar curvature of
the four-dimensional factor, which for a four-dimensional space form
is (in absolute value) $12 R^{-2}$, where $R$ is the radius of
curvature.  Therefore, in terms of the radii of curvature, the above
$4$-forms can be written as
\begin{equation*}
  F_4 = 3 R_{\AdS}^{-1} \dvol(\AdS_4)~,
\end{equation*}
and
\begin{equation*}
  F_4 = 3 R_S^{-1} \dvol(S^4)~,
\end{equation*}
respectively.

Next consider the case of the $\AdS_7\times S^4$ solution M5 brane.
The metric $g_S$ of the $4$-sphere of radius of curvature $R$ is
\begin{equation*}
  R^{-2} g_S = d\psi^2 + (\sin\psi)^2 d\Omega^2_3~,
\end{equation*}
where $d\Omega_3^2$ is the round metric on the equatorial $3$-sphere.
The corresponding volume form is then
\begin{equation*}
  \dvol(S^4) = R^4 (\sin\psi)^3 d\psi \wedge \dvol(S^3)~,
\end{equation*}
whence the $4$-form is
\begin{equation*}
  F_4 = 3 R^3 (\sin\psi)^3 d\psi \wedge \dvol(S^3)~.
\end{equation*}
Taking the Penrose limit along the geodesic with affine parameter
$u=\psi + 2\tau$, we find
\begin{equation*}
  \bar F_4 = \tfrac32 R^3 (\sin (u/2))^3 du \wedge dy^7 \wedge dy^8
  \wedge dy^9
\end{equation*}
in Rosen coordinates.  Changing to Brinkman coordinates as in
\eqref{eq:BrinkmanCoords}, we obtain
\begin{equation*}
  R^{-3} \bar F_4 = 3 dx^- \wedge dx^7 \wedge dx^8
  \wedge dx^9~,
\end{equation*}
which agrees with the expression for the $4$-form in the maximally
supersymmetric Hpp-wave solution of eleven-dimensional supergravity
\cite{KG,FOPflux}.  The same holds (after a change of variables) for
the $4$-form in the $\AdS_4\times S^7$ solution.

Finally we consider the $\AdS_5\times S^5$ solution.  In our
conventions, the self-dual $5$-form in the $\AdS_5\times S^5$ solution
is given by
\begin{equation*}
  F_5 = \half R^{-1} \left( \dvol(\AdS_5) + \dvol(S^5) \right)~,
\end{equation*}
where $R$ is the radius of curvature of both anti-de~Sitter spacetime
and the sphere.  In terms of the coordinates in which we wrote the
metrics, we have
\begin{footnotesize}
  \begin{equation*}
    F_5 = \half R^4 \left( \frac{r^3}{\sqrt{1+r^2}} (\sin\tau)^4 d\tau
      \wedge dr \wedge \dvol(S^3) + (\sin\psi)^4 d\psi \wedge \dvol(S^4)
    \right)~.
  \end{equation*}
\end{footnotesize}
Taking the Penrose limit along the geodesic parametrised by $u = \psi
+ \tau$, we obtain
\begin{footnotesize}
  \begin{equation*}
    \bar F_5 = \tfrac14 R^4 (\sin(u/2))^4 du \wedge \left( dy^1 \wedge dy^2
      \wedge dy^3 \wedge dy^4 + dy^5 \wedge dy^6 \wedge dy^7 \wedge dy^8
    \right)
  \end{equation*}
\end{footnotesize}
in Rosen coordinates.  Using \eqref{eq:BrinkmanCoords} to go to
Brinkman coordinates, we obtain
\begin{equation*}
  R^{-4} \bar F_5 = \half dx^- \wedge \left( dx^1 \wedge dx^2
  \wedge dx^3 \wedge dx^4 + dx^5 \wedge dx^6 \wedge dx^7 \wedge dx^8
  \right)~,
\end{equation*}
which, together with \eqref{eq:IIBHpp},  agrees
with the maximally supersymmetric Hpp-wave discovered in
\cite{NewIIB}.

In summary, we conclude that the maximally supersymmetric Hpp-waves of
\cite{KG,FOPflux} and \cite{NewIIB} appear as Penrose limits of
maximally supersymmetric $\AdS\times S$ spacetimes.  Moreover this is
not an accident.  In fact, it can be shown \cite{Limits} that there
are only two possible Penrose limits of these $\AdS\times S$
solutions: generically one obtains the maximally supersymmetric
Hpp-waves, but one can also get flat space for very special null
geodesics, namely those which have no velocity component tangent to
the sphere.

\subsection{Symmetries}

One intriguing feature of the maximally supersymmetric Hpp-waves is
that their symmetry (super)algebra has the same dimension as that of
the $\AdS \times S$ solutions.  It was observed in \cite{NewIIB} that
the symmetry (super)algebras of these classes of solutions are related
essentially by a contraction, and this in turn suggests a limiting
procedure relating the two classes of solutions.  This limiting
procedure is none other than the Penrose limit, as we will now show.
We will be brief, leaving the details to our forthcoming paper
\cite{Limits} which also includes a more general discussion of what
happens to isometries and supersymmetries under the Penrose limit.

Let $\xi$ be a Killing vector of the $\AdS \times S$ solution and let
us change coordinates to those considered above that are adapted to a
null geodesic.  After the change of variables \eqref{eq:diffeo}, the
vector field acquires a dependence on the scaling parameter $\Omega$.
Let us make this manifest by writing it as $\xi(\Omega)$.  For all
$\Omega>0$, $\xi(\Omega)$ is a Killing vector with respect to the
rescaled metric $g(\Omega)$; hence in the limit,
$\bar\xi=\lim_{\Omega\to0} \Omega^\Delta \xi(\Omega)$, where $\Delta$
is chosen so that the above limit exists and is non-zero, is a Killing
vector with respect to the limiting metric $\bar g$.  An argument
originally due to Geroch \cite{Geroch} and described in more detail in
the present context in \cite{Limits} shows that all Killing vectors
are preserved, although the algebraic structure is generally
contracted, due to the fact that different Killing vectors may have to
be rescaled with different values of $\Delta$.  A similar argument
shows that the Killing spinors of a supergravity solution are also
preserved.  It is important to notice, however, that the Penrose limit
might admit additional (super)symmetries not present in the original
solution.

Concretely, consider the Penrose limit of $\AdS_{p+2}\times
S^{D{-}p{-}2}$ with radii of curvature $R_{\AdS}=\rho R$ and $R_S=R$,
respectively.  The Penrose limit along the null geodesic considered
above is an Hpp-wave metric of the form \eqref{eq:CWtype} where the
matrix $A_{ij}$ is constant and has two eigenvalues with ratio $\rho$
and multiplicities $p+1$ and $D{-}p{-}3$.  Under the Penrose limit,
the $\fso(2,p+1) \oplus \fso(D{-}p{-}1)$ isometry algebra of
$\AdS_{p+2}\times S^{D{-}p{-}2}$ gets modified in the following way.
The $\fso(2,p+1)$ factor gets contracted to
$\fh(p+1)\rtimes\fso(p+1)$, where $\fh(p+1)$ is a Heisenberg algebra
with $2p+3$ generators and $\fso(p+1)$ acts on the creation and
annihilation operators in the natural way (i.e., they transform as
vectors). Similarly the $\fso(D{-}p{-}1)$ factor contracts to
$\fh(D{-}p{-}3)\rtimes\fso(D{-}p{-}3)$.  The central element in both
Heisenberg algebras coincide.  This means that there are two Killing
vectors $\xi_1(\Omega)$ and $\xi_2(\Omega)$ such that they agree to
leading order in $\Omega$ and hence coincide in the limit.  Consider
instead the linear combinations $\xi_\pm(\Omega) = \xi_1(\Omega) \pm
\xi_2(\Omega)$.  These vector fields must be rescaled differently for
their limits to exist and be non-zero: $\xi_+(\Omega)$ becomes in the
limit the common central element of the combined Heisenberg algebra
$\fh(D{-}2)$, whereas $\xi_-(\Omega)$ becomes an outer automorphism
commuting with $\fso(p+1)\oplus\fso(D{-}p{-}3)$.  In Brinkman
coordinates, $\bar\xi_\pm$ are realised as $\d/\d x^\pm$.  We see,
therefore, that the isometry algebra $\fso(2,p+1) \oplus
\fso(D{-}p{-}1)$ of $\AdS_{p+2}\times S^{D{-}p{-}2}$ contracts to a
semidirect product
\begin{equation*}
  \fh(D{-}2) \rtimes \left( \fso(p+1) \oplus \fso(D{-}p{-}3) \oplus
  \RR \right)~.
\end{equation*}
If the ratio $\rho$ of the radii of curvature is equal to $1$, there
is an additional enhancement of the isometry, and the subalgebra
$\fso(p+1)\oplus\fso(D{-}p{-}3)$ is enlarged to the full
$\fso(D{-}2)$.  This however has no counterpart in the original
metric.

In summary, the Penrose limit provides a natural explanation for the
isometry structure of the maximally supersymmetric Hpp-waves discussed
in \cite{NewIIB}.  Finding such an explanation was one of the
motivations for the present work.

\section{Worldvolume dynamics and Penrose limits}
\label{sec:dynamics}

To investigate the effect of the Penrose limit on the dynamics of
brane probes, we consider the action of a typical $p$-brane with
tension $T_p$ propagating in a $\AdS\times S$ background. Here we
shall focus on the bosonic part of brane probe actions without
worldvolume fields of Born-Infeld type. The analysis which follows can
be extended to include Born-Infeld fields and this will be presented
in \cite{Limits}.  Under these assumptions, the action of a $p$-brane
probe in a background (e.g., $\AdS \times S$) with non-vanishing
form-field potential $C$ is
\begin{equation*}
  I_p[g, C; \kappa_p]= T_p \left(\int_D d^{p+1}\sigma \sqrt{ g} +
    \int_D C\right)~,
\end{equation*}
where $T_p=1/\kappa_p$ is the $p$-brane tension and $D$ is a region of
worldvolume. The first term in the action is the usual induced volume
term and the second is a Wess--Zumino term; we do not distinguish
between the induced metric and form-gauge potential and the associated
spacetime objects $g$ and $C$, respectively.  It is clear from the
context which is which.

To implement the Penrose limit for the probe action above, we first
rescale $\kappa_p\rightarrow \Omega^{p+1}\kappa_p$, obtaining
\begin{equation*}
  I[g, C;\Omega^{p+1}\kappa_p]=I[\Omega^{-2}g, \Omega^{-(p+1)} C;
  \kappa_p]~.
\end{equation*}
Next, adopting appropriate coordinates for the Penrose limit and
performing the coordinate transformation \eqref{eq:diffeo}, we can
write
\begin{equation*}
  I[\Omega^{-2}g, \Omega^{-(p+1)} C; \kappa_p] =
  I[\Omega^{-2}g(\Omega), \Omega^{-(p+1)} C(\Omega); \kappa_p]~.
\end{equation*}
Now for $\Omega\ll 1$ we can expand the probe action in a power series
in $\Omega$ as
\begin{equation*}
  I[\Omega^{-2}g, \Omega^{-(p+1)} C; \kappa_p]=I[\bar g, \bar C;
  \kappa_p]+ O(\Omega)~,
\end{equation*}
where $(\bar g, \bar C)$ are the metric and gauge-form potential at
the Penrose limit. Now we have seen that there are two types of
Penrose limits for $\AdS\times S$ spacetimes, either the maximally
supersymmetric Hpp-wave or Minkowski spacetime. This limit takes
worldvolume solutions of probes in $\AdS\times S$ to worldvolume
solutions of probes in the Penrose limit of $\AdS\times S$, which is
either an Hpp-wave spacetime or Minkowski space, depending on the
choice of null geodesic.

A physical interpretation of the Penrose limit is as a particular
{large tension} limit of $p$-branes in a given spacetime, arising from
making the $\Omega$-dependent change of coordinates \eqref{eq:diffeo}
and the rescaling of the tension $T_p \to T_p/ \Omega ^{p+1}$. The
theory can then be developed as a perturbation series in $\Omega$, and
there will be a different perturbation series for different choices of
null geodesic.  Taking the limit $\Omega \to 0$ then reduces the brane
action to that of a brane in a spacetime which is the Penrose limit of
the original spacetime, and this will depend on the choice of null
geodesic \cite{Limits}.  In this limit $\kappa_p\to 0$, so it is a
weak coupling limit from the persepctive of the $p$-brane.  For branes
in $\AdS\times S$, the limit arising in this way is Minkowski
spacetime or the maximally supersymmetric Hpp-wave, depending on the
choice of geodesic.

In particular we find that the IIB string in $\AdS_5\times S^5$ has two
large tension limits of this kind: the string in ten-dimensional
Minkowski spacetime and the string in the Hpp-wave solution found in
\cite{NewIIB}. It is worth mentioning that in both of these limits the
IIB superstring Green--Schwarz action does not exhibit any
interactions after gauge fixing (see \cite{MetsaevIIB} for the case of
the Hpp-wave) despite the presence of Ramond-Ramond field and can be
quantised exactly.

\section*{Acknowledgments}

This work was initiated while three of us (MB,JMF,CMH) were
participating in the programme \emph{Mathematical Aspects of String
  Theory} at the Erwin Schrödinger Institute in Vienna, whom we would
like to thank for support.  The research of MB is partially supported
by EC contract CT-2000-00148.  JMF is a member of EDGE, Research
Training Network HPRN-CT-2000-00101, supported by The European Human
Potential Programme.  GP is supported by a University Research
Fellowship from the Royal Society.  This work is partially supported
by SPG grant PPA/G/S/1998/00613.  In addition, JMF would like to
acknowledge a travel grant from PPARC.

%

\begin{thebibliography}{10}

\bibitem{NewIIB}
M~Blau, JM~Figueroa-O'Farrill, CM~Hull, and G~Papadopoulos, \emph{A new
  maximally supersymmetric background of type {IIB} superstring theory},
  \texttt{arXiv:hep-th/0110242}.

\bibitem{Limits}
\bysame, \emph{Penrose limits in string theory}, in preparation.

\bibitem{CahenWallach}
M~Cahen and N~Wallach, \emph{Lorentzian symmetric spaces}, Bull. Am. Math. Soc.
  \textbf{76} (1970), 585--591.

\bibitem{CKG}
PT~Chru{\'s}ciel and J~Kowalski-Glikman, \emph{The isometry group and {K}illing
  spinors for the pp wave space-time in ${D}=11$ supergravity}, Phys. Lett.
  \textbf{149B} (1984), 107--110.

\bibitem{FOPflux}
JM~Figueroa-O'Farrill and G~Papadopoulos, \emph{Homogeneous fluxes, branes and
  a maximally supersymmetric solution of {M}-theory}, J. High Energy Phys.
  \textbf{06} (2001), 036, \texttt{arXiv:hep-th/0105308}.

\bibitem{Geroch}
R~Geroch, \emph{Limits of spacetimes}, Comm. Math. Phys. \textbf{13} (1969),
  180--193.

\bibitem{GuevenPlaneWave}
R~Güven, \emph{Plane wave limits and {T}-duality}, Phys. Lett. \textbf{B482}
  (2000), 255--263, \texttt{arXiv:hep-th/0005061}.

\bibitem{KG}
J~Kowalski-Glikman, \emph{Vacuum states in supersymmetric {K}aluza-{K}lein
  theory}, Phys. Lett. \textbf{134B} (1984), 194--196.

\bibitem{Meessen}
P~Meessen, \emph{A small note on pp-wave vacua in 6 and 5 dimensions},
  \texttt{arXiv:hep-th/0111031}.

\bibitem{MetsaevIIB}
RR~Metsaev, \emph{Type {IIB} {G}reen--{S}chwarz superstring in plane wave
  {R}amond-{R}amond background}, \texttt{arXiv:hep-th/0112044}.

\bibitem{ORSPL}
DI~Olive, E~Rabinovici, and A~Schwimmer, \emph{A class of string backgrounds as
  a semiclassical limit of {WZW} models}, Phys. Lett. \textbf{B321} (1994),
  361--364, \texttt{arXiv:hep-th/9311081}.

\bibitem{PenrosePlaneWave}
R~Penrose, \emph{Any space-time has a plane wave as a limit}, Differential
  geometry and relativity, Reidel, Dordrecht, 1976, pp.~271--275.

\bibitem{SfetsosPL}
K~Sfetsos, \emph{Gauging a nonsemisimple {WZW} model}, Phys. Lett.
  \textbf{B324} (1994), 335--344, \texttt{arXiv:hep-th/9311010}.

\bibitem{SfetsosTseytlinPL}
K~Sfetsos and AA~Tseytlin, \emph{Four-dimensional plane wave string solutions
  with coset {CFT} description}, Nucl. Phys. \textbf{B427} (1994), 245--272,
  \texttt{arXiv:hep-th/9404063}.

\bibitem{TseytlinExact}
AA~Tseytlin, \emph{Exact string solutions and duality},
  \texttt{arXiv:hep-th/9407099}.

\end{thebibliography}

\providecommand{\bysame}{\leavevmode\hbox to3em{\hrulefill}\thinspace}

\end{document}